# Comment on "Spin-Orbit Logic with Magnetoelectric Nodes: A Scalable Charge Mediated Nonvolatile Spintronic Logic"  (arXiv:1512.05428)

D. C. Ralph

Laboratory of Atomic and Solid State Physics, Cornell University, Ithaca, NY 14853

Kavli Institute at Cornell for Nanoscale Science, Ithaca, NY 14853

Researchers from Intel Corporation recently proposed "Magneto-Electric Spin Orbit (MESO)" logic as a new strategy for beyond-CMOS electronics [1]. The Intel researchers project that this concept has the potential to reduce the switching energy per bit "to near thermodynamic limits for GHz logic (100 kT switching at 100 ps delay)", *i.e.*, to near 0.4 aJ. Here I point out that the switching energy stated in ref. [1] is incorrect, because the paper neglects a large energy cost associated with Ohmic dissipation that is unavoidable within the MESO scheme. Using optimistic parameters, the true minimum switching energy per bit within the MESO approach is at least 150 fJ, or more than 300,000 times greater than the value stated in ref. [1]. Given this large energy cost, the MESO concept is not a viable strategy for beyond-CMOS logic.

The existence of a large energy cost due to Ohmic dissipation in the MESO logic scheme can be understood from an elementary analysis of the circuit diagram in Fig. 1 (identical to Fig. 3B of ref. [1] with some labels added). Switching is achieved by first passing a spin-polarized charge current $I_{in}$ through a ferromagnetic injector (from point A through the impedance labeled as $R_{FMI}$, then through point B to ground), which generates a charge current within a spin-to-charge converter. Depending on the spin orientation of $I_{in}$, the sign of the output charge current may be positive or negative. The output charge current from the spin-to-charge converter then produces charge accumulation on the capacitor labeled as $C_{ME}$, thereby generating an electric field to drive magneto-electric switching of a magnetic element. In estimating the total energy cost of this process, ref. [1] asserted that the dominant energy cost is associated with the electrostatic energy stored on the capacitor $C_{ME}$. However, by the nature of the materials that must be used within the spin-to-charge converter (a metal layer in contact to a material with strong spin-orbit coupling), it necessarily functions as a current source with a small internal resistance, $R_{SCC}$. The value of $R_{SCC}$ is estimated in ref. [1] to be at most approximately 10 Ω. Reference [1] did not take into account that, like any current source with an internal resistance much less than the load, more power will be dissipated within the internal resistance than delivered to the output, in this case much more. The small internal resistance $R_{SCC}$ effectively lies in parallel with the magneto-electric capacitor $C_{ME}$ (given that $R_{SCC}$ provides a low-impedance path from point C in Fig. 1 to ground). Therefore, for the full time that $C_{ME}$ is biased with a steady voltage to drive magnetic switching (assumed in [1] to be a 100 mV bias for 100 ps), the voltage that is present across $C_{ME}$ must also drop across the resistance $R_{SCC}$, where it will generate a large current flow and a large amount of Ohmic dissipation. This current flow must circulate within the spin-to-charge converter for the full time required to drive magnetic switching; if the spin-to-charge converter is not energized, the magneto-electric capacitor $C_{ME}$ will discharge through $R_{SCC}$ and switching will not occur.

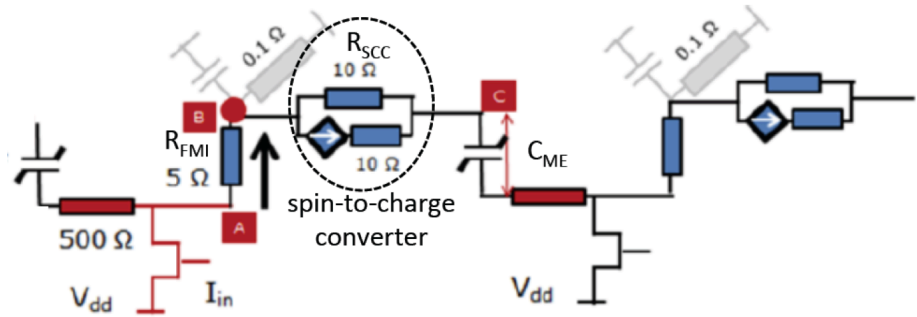

Fig. 1.  Circuit diagram of MESO logic, from [1] with some added labels.

Assuming the optimistic parameters stated in ref. [1]: that a voltage of 100 mV must be applied to the magneto-electric capacitor for 100 ps to drive switching and that $R_{SCC} = 10\ \Omega$, the energy cost due to Ohmic dissipation in $R_{SCC}$ for each switching event is (100 ps) (100 mV)$^2$/(10 $\Omega$) = 100 fJ.

There will be additional Ohmic dissipation due to current flow through the ferromagnetic injector, $R_{FMI}$. Assuming, generously, that the spin-polarized charge current ($I_{in}$) is 100% polarized and that the spin-to-charge converter has unity efficiency, the magnitude of $I_{in}$ must be at least approximately the same as the current generated within the spin-to-charge converter, which for the parameters stated above is $I_{SCC} = $ (100 mV)/(10 $\Omega$) = 10 mA [2]. Using the value $R_{FMI} = 5\ \Omega$ projected in ref. [1], the minimum energy dissipated in $R_{FMI}$ during a 100 ps switching process is then (100 ps) (10 mA)$^2$ (5 $\Omega$) = 50 fJ. Together, the sum of the Ohmic energy loss per switching event in $R_{SCC}$ and $R_{FMI}$ is therefore at least 150 fJ, the value I stated above.

1. S. Manipatruni, D. E. Nikonov, and I. A. Young, "Spin-Orbit Logic with Magnetoelectric Nodes: A Scalable Charge Mediated Nonvolatile Spintronic Logic," arXiv:1512.05428.

2. It should also be noted that 10 mA is a huge current per bit, and this by itself would prevent the scaling of MESO devices to the small dimensions needed for a beyond-CMOS logic technology.